\documentclass[prd,aps,
floats,letterpaper,floatfix,
groupedaddress,superscriptaddress
,nofootinbib
,eqsecnum
]{revtex4}
\usepackage{amsmath}
\usepackage{amsfonts}
\usepackage{bm}
\usepackage{graphicx}
\usepackage[usenames]{color}
  \newcommand{\apjl}{Astrophys. J. Lett.}

   \newcommand{\lrr}{Living Rev. Relativ.}

\def\v1v2{{\bf v}_1 \cdot {\bf v}_2}

\allowdisplaybreaks

\begin{document}

\title{Modified geodesic equations of motion for compact bodies in alternative theories gravity}

\author{
Fatemeh Taherasghari} \email{ftaherasghari@ufl.edu}
\affiliation{Department of Physics, University of Florida, Gainesville, Florida 32611, USA}

\author{
Clifford M.~Will} \email{cmw@phys.ufl.edu}
\affiliation{Department of Physics, University of Florida, Gainesville, Florida 32611, USA}
\affiliation{ 
GReCO, Institut d'Astrophysique de Paris, CNRS, Sorbonne Universit\'e, 98 bis Bd.\ Arago, 75014 Paris, France}

\date{\today}

\begin{abstract}
We derive exact, modified geodesic equations for a system of non-spinning, self-gravitating interacting bodies in a class of alternative theories of gravity to general relativity.  We use a prescription proposed by Eardley for incorporating the effects of self-gravity within gravitationally bound bodies, in which their masses may depend on invariant quantities constructed from the auxiliary scalar, vector or tensor fields introduced by such theories, evaluated in the vicinity of each body.  The forms of the equations are independent of the field equations of the chosen theory.  In the case where the masses are strictly constant, the equations reduce to the conventional geodesic equations of general relativity.  These equations may be useful tools for deriving equations of motion for compact bodies to high post-Newtonian orders in alternative theories of gravity.
\end{abstract}

\maketitle

\section{Introduction}
\label{sec:intro}
The motion of self-gravitating bodies in relativistic gravitational theories such as general relativity is a problem of both theoretical and observational interest.  From a theoretical point of view, the problem raises fundamental questions of principle, such as how to separate internal dynamics from the overall motion of the bodies \cite{Damour300}, how to define usable (if not entirely rigorous) ``centers of mass'', and the efficacy of ``point particles'' as a tool for treating ``small'', well separated bodies.  From an observational point of view the problem has been important ever since Nordtvedt's 1967 discovery that, in a broad class of alternative theories of gravity, the acceleration of self-gravitating bodies in an external gravitational field could depend on their gravitational binding energy \cite{1968PhRv..169.1017N}, thereby violating the principle of equivalence (today called the Strong Equivalence Principle), and providing an important test of gravitational theories via studies of the Earth-Moon orbit in the field of the Sun using lunar laser ranging \cite{tegp2}. 

The 1974 discovery of the Binary Pulsar B1913+16 \cite{1975ApJ...195L..51H} made the problem more acute, because the self-gravity of the neutron stars in the system involved {\em strong} gravitational fields, in contrast to the weak fields that characterize the Earth and the Moon's internal structure.  And indeed, in an influential 1976 paper, Ehlers et al.\   \cite{1976ApJ...208L..77E} criticized the entire state of work on the problem of motion as it stood at the time, questioning among other things the use of delta functions to treat small well-separated bodies in general relativity.   That paper, together with the emerging idea that the inspiral of binary neutron stars or black holes might be an important source of gravitational waves for the nascent laser interferometric detectors, motivated a major effort by many research groups to put the ``problem of motion and radiation'' on a firmer theoretical foundation (see \cite{2014LRR....17....2B,2007tste.book.....K} for reviews).  

One aspect of that program was the development of sophisticated regularization techniques (Hadamard regularization, dimensional regularization) to handle the infinities that go hand in hand with the use of delta functions to describe the bodies.  At the same time, it was shown that, at least through second post-Newtonian (2PN) order in general relativity, for a system of small but finite-size fluid balls, taking into account their internal gravitational structure, isolating terms that scale as inverse powers of the size of each body, and making a universal renormalization of their masses through 2PN order to eliminate all such seemingly singular terms, the equations of motion were identical to those obtained using delta-function sources with appropriate regularizations  \cite{2007PhRvD..75l4025M}
(see also \cite{1983SvAL....9..230G}).  This gave rise to the widely accepted assumption that the motion of non-spinning bodies in GR would be independent of their internal structure (up to tidal interactions), and would thus be amenable to treatment using delta function sources, at least to some high post-Newtonian order (Damour \cite{Damour300} estimated that it might fail at 5PN order).

What about alternative theories of gravity?  Because many alternative theories introduce auxiliary long-range gravitational fields (scalar, vector, tensor, $\dots$) in addition to the metric, the structure of self-gravitating bodies may depend on the asymptotic values of these fields induced by cosmology or even by nearby bodies in the system  \cite{1972ApJ...177..757W,tegp2}, and in particular, their masses could depend on these values.  But if a body's mass varies as it moves in the field of other bodies, its acceleration must be modified appropriately in order to preserve energy conservation (see the classic cyclic {\em gedanken} experiment arguments by Dicke \cite{1964rgt..book.....Dicke}, Nordtvedt \cite{1975PhRvD..11..245N}, and Haugan \cite{1979AnPhy.118..156H}).  These structure-dependent modifications of the motion lead to violations of the Strong Equivalence Principle, as exemplified in the Nordtvedt effect.

In 1975, while studying the implications of the binary pulsar discovery for the alternative scalar-tensor theory of gravity, Eardley \cite{1975ApJ...196L..59E} devised a prescription that incorporated violations of the SEP, permitted strong internal gravitational fields, yet retaining the simplifying aspects of using delta-functions for the sources.  He proposed to write the matter action of a system of non-spinning, self-gravitating ``point'' particles in the form
\begin{equation}
S_M = - \sum_A \int m_A (\phi) d\tau \,,
\label{eq:action0}
\end{equation}
where $\phi$ is the scalar field evaluated at the location of each particle, and $\tau$ is proper time along the world line of each particle.   The resulting scalar-tensor field equations 
and equations of motion then depended on what Eardley called ``sensitivities'' of each particle's mass to variations of the scalar field, defined by 
\begin{equation}
s_A \equiv \left ( \frac{d \ln m_A (\phi)}{d \ln \phi} \right ) \,.
\end{equation}
With the use of  Eardley's prescription combined with estimates of sensitivities for neutron stars using various equations of state, bounds have been placed on  alternative theories of gravity using binary pulsar data, including Rosen's bimetric theory \cite{1977ApJ...212L..91W},
massless and massive scalar-tensor theory \cite{1989ApJ...346..366W,1996PhRvD..54.1474D,1998PhRvD..58d2001D,2012PhRvD..85f4041A,2012MNRAS.423.3328F}, and
Einstein-{\AE}ther theory \cite{2014PhRvL.112p1101Y,2014PhRvD..89h4067Y}.
It has  been used to obtain equations of motion for compact bodies in scalar-tensor gravity  to 3PN order  \cite{2013PhRvD..87h4070M,2018PhRvD..98d4004B}, the gravitational waveforms to 2PN order for tensor waves and to 1.5PN order for scalar waves \cite{2015PhRvD..91h4027L,2014PhRvD..89h4014L},  and the energy flux  to 1.5PN order
\cite{2022arXiv220110924B}.  It has also been used to obtain the 1PN equations of motion in Einstein-{\AE}ther theory  as well as the leading radiative effects   \cite{2007PhRvD..76h4033F,2014PhRvD..89h4067Y}.

In an effort to investigate the range of validity of Eardley's prescription, Gralla \cite{2010PhRvD..81h4060G,2013PhRvD..87j4020G} developed a more general and rigorous theory of the motion of ``small'' bodies characterized by such parameters as mass, spin and charge in an environment of external fields, and
 argued that Eardley's prescription is a special case of that general framework.
 
In this paper, we explore the {\em practical} implications of the Eardley prescription for the equations of motion of compact bodies in alternative theories of gravity.   
We consider a system of ``point'' masses, where each mass is a scalar quantity that depends one or more invariant quantities $\psi$ constructed from the auxiliary gravitational fields of the theory, the spacetime metric, and the four-velocity of the mass, evaluated ``at'' the location of each body.  By ``at'', we mean more precisely that the quantities are evaluated at the boundary of a region surrounding each body that is sufficiently large that the metric of the body is approximately asymptotically flat, yet sufficiently small compared to the interbody distances that variations of the quantities across the region may be ignored.  In other words, we ignore all tidal effects arising from the actual finite size of real bodies.   We also ignore spin effects.

The simplest example, of course, is a scalar field, whereby the mass is a function of $\phi$; in scalar-tensor theories $\phi$ is directly related to the local value of Newton's constant, which clearly influences the body's internal structure and gravitational binding energy. For a timelike vector auxiliary field $K^\mu$, the mass might depend on $K^{\hat{0}}$, the time component of $K^\mu$ as measured in the quasi-local comoving inertial frame of the body (since the body is assumed to be ``spherical'', it is hard to see how the mass could depend on a local spatial component $K^j$).  The invariant quantity whose value is $K^{\hat{0}}$ in the comoving frame is $- u_A^\mu K^\nu g_{\mu\nu}$, where $u_A^\mu$ is the four-velocity of body $A$.  The mass could also depend on the magnitude $K^\mu K_\mu$.  
For this paper, we will consider a specific set of invariant quantities arising from the auxiliary fields $\phi$, $K^\mu$ and a tensor field $C^{\mu\nu}$, divided into two categories:  (A): invariants constructed from the fields only, and (B): invariants involving the bodies' four-velocities:
\begin{align}
\textrm{A:} \quad& \phi \,, \quad  -K^\mu K_\mu \,,
\quad  -{C^\mu}_\mu \,, \quad C^{\mu\nu} C_{\mu\nu} \,,  \quad C^{\mu\nu} C_{\nu\mu} \,;
\nonumber \\
\textrm{B:} \quad & - K^\mu u_{A \mu} \,, \quad 
  C^{\mu\nu} u_{A\mu} u_{A\nu}\,, \quad  C^{\alpha\mu} {C_\alpha}^{\nu} u_{A\mu} u_{A\nu}  \,, 
  \nonumber \\ 
&\qquad  C^{\alpha\mu} {C^{\nu}}_\alpha u_{A\mu} u_{A\nu} \,.
\label{eq:invariants}
\end{align}
We do not assume any {\em a priori} symmetry for the tensor field $C^{\mu\nu}$.  The negative signs are purely conventional.  If we assume that the vector $K^\mu$ and the tensor $C^{\mu\nu}$ are ``timelike'' in the sense that in the rest frame of a homogeneous isotropic universe, they have no spatial components (apart from a possible spatial $\delta^{ij}$) and have positive $K^0$ or $C^{00}$, then the invariants will be positive in the vicinity of the bodies.  

These are the simplest possible couplings involving scalar, vector or tensor fields.  If a theory of gravity should involve higher-rank or more exotic fields, or couplings beyond quadratic order, the methods described here should be generalizable to more complicated cases.  Starting from a matter action of the form of Eq.\ (\ref{eq:action0}) we derive a set of exact, ``modified geodesic equations'' for each invariant, in two ways, in Sec.\ \ref{sec:pointmass} by applying the standard calculus of variations to the action, and then in Sec.\ \ref{sec:bianchi}, by using the general covariance of the action to obtain a modified matter ``Bianchi identity'' analogous to the general relativistic vanishing of the divergence of the energy-momentum tensor.  Both methods yield the same modified geodesic equations.  Section \ref{sec:conclusions} makes concluding remarks.

\section{Compact body equation of motion from a point-mass action}
\label{sec:pointmass}

Applying Eardley's prescription, we write down the action for a system of self-gravitating point masses,
\begin{equation}
S_M = -\sum_A  \int_P^Q \, m_A (\psi) (-g_{\mu\nu} v^\mu v^\nu )^{1/2} d\lambda \,,
\label{eq:actionpoint}
\end{equation}
 where $\lambda$ is a timelike integration variable along the worldline of each particle, $v^\mu = dx^\mu/d\lambda$, and $\psi$ stands for one of the invariant expressions in (\ref{eq:invariants}). All variables in $S_M$ are to be evaluated at the location of body A.   In theories where there might be more than one auxiliary field, or where more than one invariant might be involved in determining the mass of body A, there might be several $\psi$'s: $\psi_1$, $\psi_2$ and so on, but to keep the notation simple we will assume that only one field and invariant is involved.  Generalization to multiple $\psi$'s is straightforward.  
 
We now vary the path of each particle by $\delta x_A^\alpha (\lambda)$, with $\delta x_A^\alpha (P) = \delta x_A^\alpha (Q ) = 0$.  Noting that $(-g_{\mu\nu} v^\mu v^\nu )^{1/2} = d\tau_A/d\lambda$, where $\tau_A$ is proper time along the worldline of body A,  and expanding about the unperturbed worldline, we obtain, to first order in the deviations $\delta x^\alpha$,
\begin{align}
\frac{d\tau_A}{d\lambda} &= \left ( \frac{d\tau_A}{d\lambda} \right )_0 
- \frac{1}{2} g_{\mu\nu , \alpha} u_A^\mu u_A^\nu \left ( \frac{d\tau_A}{d\lambda} \right )_0 \delta x_A^\alpha 
\nonumber \\
& \qquad
- g_{\alpha\nu} u_A^\nu \frac{d}{d\lambda} \delta x_A^\alpha \,,
\nonumber \\
m_A (\psi) &= m_{A}  + m'_A \left ( \frac{\partial \psi}{\partial u_A^\mu} \delta u_A^\mu
+  \frac{\partial \psi}{\partial g_{\mu\nu}} \delta g_{\mu\nu} +  \frac{\partial \psi}{\partial \phi} \delta \phi
\right . 
\nonumber \\ 
&
\left . \qquad 
+ \frac{\partial \psi}{\partial K^\mu} \delta K^\mu +
 \frac{\partial \psi}{\partial C^{\mu\nu}} \delta C^{\mu\nu} \right ) \,,
\end{align}
where $m_A \equiv m_A (\psi_0)$ and $m'_A \equiv \partial m_A /\partial \psi_0$, where the subscript $0$ denotes evaluation along the unperturbed paths.  Note that
\begin{align}
\delta u_A^\mu &= \left (\frac{d\lambda}{d\tau_A} \right )_0   \left ( \delta^\mu_\alpha + u_A^\mu u_{A\alpha} \right ) \frac{d}{d\lambda} \delta x_A^\alpha
 \nonumber \\
 & \qquad
 +\frac{1}{2} g_{\beta\gamma, \alpha}  u_A^\beta u_A^\gamma u_A^\mu \delta x_A^\alpha \,,
 \nonumber \\
 \delta g_{\mu\nu} &= g_{\mu\nu,\alpha} \delta x_A^\alpha 
 \,,
 \nonumber \\
 \delta \phi &= \phi_{,\alpha} \delta x_A^\alpha 
 \,,
 \nonumber \\
\delta K^\mu &= {K^\mu}_{,\alpha} \delta x_A^\alpha
 \,,
 \nonumber \\
\delta C^{\mu\nu} & = {C^{\mu\nu}}_{,\alpha} \delta x_A^\alpha \,. 
 \end{align}
Using these results to expand the action $S$ to first order in deviations of the worldline, integrating by parts to eliminate $d\delta x^\alpha /d\lambda$, and demanding that $\delta S = 0$ for arbitrary variations $\delta x_A^\alpha$, we obtain the equation of motion for each body A,
\begin{align}
u_A^\nu \nabla_\nu &\left [ m_A u_{A\alpha} - m'_A \frac{\partial \psi}{\partial u_A^\mu}
\left ( \delta^\mu_\alpha + u_A^\mu u_{A\alpha} \right ) \right ]
\nonumber 
\\
&\quad  = - m'_A \left [ \frac{\partial \psi}{\partial \phi} \phi_{,\alpha} 
+ \frac{\partial \psi}{\partial g_{\mu\nu}} g_{\mu\nu,\alpha}
-  \frac{\partial \psi}{\partial u_A^\mu} {\Gamma^\mu}_{\beta\alpha} u^\beta 
\right .
\nonumber \\
& \left . 
\qquad \quad
+  \frac{\partial \psi}{\partial K^\mu}  {K^\mu}_{,\alpha}
+ \frac{\partial \psi}{\partial C^{\mu\nu}} C^{\mu\nu}_{,\alpha} \right ] \,,
\label{eq:eom1}
\end{align}
where $\nabla_\nu$ is the covariant derivative with respect to the metric.

For the simple case of a scalar field, $\psi = \phi$, and the modified geodesic equation becomes 
\begin{equation}
u_A^\nu \nabla_\nu \left ( m_A (\phi) u_{A\alpha} \right ) = -m'_A  \phi_{,\alpha} \,,
\end{equation}
which is the classic scalar-tensor result in Eardley's original paper (Eq.\ (5) of 
\cite{1975ApJ...196L..59E}).

For the couplings in category (A), $\partial \psi/\partial u_A^\mu = 0$.  Furthermore, it is easy to see that the derivative $\partial \psi/\partial g_{\mu\nu}$ will generate terms involving either $K^\mu$ or $C^{\mu\nu}$ that combine with the remaining terms on the right-hand side of Eq.\ (\ref{eq:eom1}) to produce a derivative of the original invariant.  In other words for {\em all} couplings in category (A), the modified geodesic equation has the form 
\begin{equation}
u_A^\nu \nabla_\nu \left ( m_A (\psi) u_{A\alpha} \right ) = - m'_A  \psi_{,\alpha} \,,
\label{eq:modgeodesicA}
\end{equation}
where $\psi =  -K^\mu K_\mu \,,\,
 -{C^\mu}_\mu \,,\, C^{\mu\nu} C_{\mu\nu} \,,$ or  $ C^{\mu\nu} C_{\nu\mu}$.
 
 For the couplings in category (B), $\partial \psi/\partial u_A^\mu \ne 0$, and we must consider each case in turn.   For $\psi = - K^\mu u_{A\mu}$, we have that 
 $\partial \psi/\partial u_A^\mu = - K_\mu$, $\partial \psi/\partial K^\mu = - u_\mu$, and
 $\partial \psi/\partial g_{\mu\nu} = - u^{(\mu} K^{\nu )}$.  It turns out that the three terms on the  right-hand side of Eq.\ (\ref{eq:eom1}) that involve these derivatives combine to produce $- u_{A\mu} \nabla_\alpha {K^\mu}$.  Thus, for $\psi = - K^\mu u_{A\mu}$, the modified geodesic equation has the form
 \begin{align}
u_A^\nu \nabla_\nu &\left [ m_A u_{A\alpha} + m'_A K^\mu \left (g_{\mu\alpha} + u_{A\mu} u_{A\alpha} 
\right ) \right ] 
\nonumber \\
&\qquad
= m'_A u_{A\mu} \nabla_\alpha K^\mu  \,.
\label{eq:case1}
\end{align}
At 1PN order, this agrees with the modified geodesic equation used in Einstein-{\AE}ther theory for compact bodies (see Eq.\  (21) of \cite{2007PhRvD..76h4033F}).  Note that in Einstein-{\AE}ther theory, there is no coupling to $K^\mu K_\mu$, because that quantity is constrained to be equal to $-1$.

For the tensor invariant $\psi = C^{\mu\nu} u_{A\mu} u_{A\nu}$, we have that 
$\partial \psi/\partial u_A^\mu = 2 C^{(\beta\gamma)} u_{A\gamma} g_{\mu\beta}$,
$\partial \psi/\partial C^{\mu\nu} = u_{A\mu} u_{A\nu}$, and
$\partial \psi/\partial g_{\mu\nu} =  \left (C^{(\mu\delta)} u_A^\nu + C^{(\nu\delta)} u_A^\mu \right ) u_{A\delta}$.   Once again, the three terms on the  right-hand side of Eq.\ (\ref{eq:eom1}) that involve these derivatives combine to produce $-u_{A\mu} u_{A\nu} \nabla_\gamma C^{\mu\nu}$.  Thus, for $\psi = C^{\mu\nu} u_{A\mu} u_{A\nu}$ the modified geodesic equation has the form
 \begin{align}
u_A^\nu \nabla_\nu &\left [ m_A u_{A\alpha} - 2m'_A C^{(\beta\gamma)} u_{A\gamma} \left (g_{\alpha\beta} + u_{A\beta} u_{A\alpha} 
\right ) \right ] 
\nonumber \\
&\qquad
=- m'_A u_{A\beta} u_{A\gamma} \nabla_\alpha C^{\beta\gamma}  \,.
\label{eq:case2}
\end{align}

For the remaining tensor invariants
$C^{\beta\mu} {C_\beta}^{\nu} u_{A\mu} u_{A\nu} $ and 
$C^{\beta\mu} {C^{\nu}}_\beta u_{A\mu} u_{A\nu}$, the calculation proceeds in a parallel manner; for the first invariant, the result is
\begin{align}
u_A^\nu  \nabla_\nu & \left [ m_A u_{A\alpha} - 2m'_A C^{\delta\beta} {C_\delta}^\gamma u_{A\gamma} \left (g_{\alpha\beta} + u_{A\alpha}  u_{A\beta} 
\right ) \right ] 
\nonumber \\
&\qquad
=- m'_A u_{A\beta} u_{A\gamma} \nabla_\alpha  \left (C^{\delta\beta}  
{C_\delta}^\gamma \right )\,;
\label{eq:case3}
\end{align}
for the second invariant the result is the same, but with ${C^\gamma}_\delta$ replacing ${C_\delta}^\gamma$.

\section{Bianchi identities and compact bodies}
\label{sec:bianchi}

An alternative approach to obtaining equations of motion for compact bodies is based  on what are called ``Bianchi identities'' for matter (not to be confused with the Bianchi identities associated with the Riemann and Ricci tensors).  We begin by reviewing these identities for conventional metric theories of gravity constructed from a covariant action.
In the matter sector, the action is assumed to depend {\em only}  on the variables $q_A$ of the non-gravitational fields, and on the spacetime metric $g_{\mu\nu}$, in other words
\begin{equation}
S_M =  \int \sqrt{-g} \,{\cal L}_M d^4x  \,,
\label{eq:action}
\end{equation}
where
\begin{equation}
{\cal L}_M = {\cal L}_M (q_A, q_{A,\mu}, g_{\mu\nu}, g_{\mu\nu,\beta}) \,,
\label{eq:Lagrangian}
\end{equation}
where $q_A$ and $ q_{A,\mu} $ are the nongravitational fields and their first partial derivatives, and $g_{\mu\nu}$ and $g_{\mu\nu,\beta}$ are the metric and its derivatives.
 (The extension to second and higher
derivatives is straightforward). This assumption is sometimes called ``universal coupling'', a consequence of the Einstein Equivalence Principle.  The action principle $\delta S_M =0$ is covariant,
thus, under a coordinate transformation, ${\cal L}_M$ must be unchanged in functional form, modulo a divergence (see  \cite{1962gait.book.....T,1973PhRvD...7.3563T} for discussion).
Consider the infinitesimal coordinate transformation
\begin{equation}
x^\mu \to x^\mu + \xi^\mu \,.
\end{equation}
Then the metric changes according to
\begin{equation}
\delta g_{\mu\nu} = - g_{\mu\alpha} {\xi^\alpha}_{,\nu} - g_{\nu\alpha} {\xi^\alpha}_{,\mu}
- g_{\mu\nu,\alpha} \xi^\alpha \,.
\label{eq3:changeg}
\end{equation}
Assume that  the matter and nongravitational field variables change according
to
\begin{align}
\delta q_A = d^\mu_{A\nu} {\xi^\nu}_{,\mu} - q_{A,\nu} \xi^\nu \,,
\label{eq3:changeq}
\end{align}
where $d^\mu_{A\nu}$ are functions of $x^\alpha$ and encode the tensorial transformation properties of $q_A$; the second term accounts for the spacetime variation of $q_A$.  Under this transformation, $\sqrt{-g}{\cal L}_M$ changes by
\begin{widetext}
\begin{equation}
\sum_A \left ( \frac{\partial \sqrt{-g}{\cal L}_M}{\partial q_A} \delta q_A + 
\frac{\partial \sqrt{-g}{\cal L}_M}{\partial q_{A,\mu}} \delta q_{A,\mu} \right )
+  \frac{\partial \sqrt{-g}{\cal L}_M}{\partial g_{\mu\nu}} \delta g_{\mu\nu} + 
\frac{\partial \sqrt{-g}{\cal L}_M}{\partial g_{\mu\nu,\beta}} \delta g_{\mu\nu,\beta} \,.
\end{equation}
Substituting Eqs.~(\ref{eq3:changeg}) and (\ref{eq3:changeq}), integrating by parts, dropping
divergence terms, and demanding that ${\cal L}_M$ be unchanged for arbitrary
functions $\xi^\alpha$ yields the ``Bianchi identities''
\begin{align}
 \left [g_{\mu\alpha} (-g)^{1/2} T^{\mu\nu} \right ]_{,\nu}
 -\frac{1}{2} g_{\mu\nu,\alpha} (-g)^{1/2} T^{\mu\nu} 
-\sum_A &\left [ q_{A,\alpha} \frac{\delta\sqrt{-g} {\cal L}_M}{\delta q_A} + 
\left ( d^\mu_{A\alpha} \frac{\delta \sqrt{-g}{\cal L}_M}{\delta q_A} \right )_{,\mu} \right ]
 = 0 \,,
\label{eq3:bianchi}
\end{align}
where $\delta \sqrt{-g}{\cal L}_M/\delta q_A$ is the ``variational'' derivative of 
${\cal L}_M$ defined, for any variable $q_A$, by
\begin{equation}
\frac{\delta \sqrt{-g}{\cal L}_M}{\delta q_A} \equiv  \frac{\partial \sqrt{-g}{\cal L}_M}{\partial q_A} - \frac{\partial}{\partial x^\mu} \left ( \frac{\partial \sqrt{-g}{\cal L}_M}{\partial q_{A,\mu}} \right ) \,,
\end{equation}
and $T^{\mu\nu}$ is the ``energy-momentum tensor'', defined by
\index{Energy-momentum tensor} 
\begin{equation}
T^{\mu\nu} \equiv \frac{2}{\sqrt{-g}} \frac{\delta \sqrt{-g}{\cal L}_M}{\delta g_{\mu\nu}} \,.
\label{eq3:Tmunudefined}
\end{equation}
Using the fact that
$\partial_\mu (-g)^{1/2}  = (-g)^{1/2} \Gamma^\alpha_{\mu\alpha}$,
we can rewrite Eq.~(\ref{eq3:bianchi}) in the form
\begin{equation}
\nabla_\nu T^\nu_{\alpha} = (-g)^{-1/2} \sum_A \left [ q_{A,\alpha} \frac{\delta \sqrt{-g}{\cal L}_M}{\delta q_A} + 
\left ( d^\mu_{A\alpha} \frac{\delta \sqrt{-g}{\cal L}_M}{\delta q_A} \right )_{,\mu} \right ] \,.
\label{eq3:bianchi2}
\end{equation}
However, the nongravitational field equations (eg. Maxwell's equations, electroweak field equations, quantum chromodynamics) and equations of motion
are obtained by setting the variational derivative of ${\cal L}_M$ with respect to
each matter field variable $q_A$ equal to zero, i.e.,
\begin{equation}
\delta \sqrt{-g}{\cal L}_M/\delta q_A = 0 \,,
\label{eq:nongrav}
\end{equation}
which by Eq.~(\ref{eq3:bianchi2}) is equivalent to
$\nabla_\nu T^\nu_{\alpha} = 0$.
Thus, the vanishing of the divergence of the energy-momentum tensor $T^{\mu\nu}$ {\em is a
consequence of the nongravitational equations of motion}, and is valid in any metric theory of gravity\footnote{Numerous GR textbooks make the statement that the Riemannian Bianchi identity satisfied by the Einstein tensor {\em implies} the vanishing divergence of the energy momentum tensor, and indeed Einstein felt that this was a crucial feature of his new theory.  But it is not true (or at best is irrelevant).  The vanishing of the divergence of $T^{\mu\nu}$ holds in {\em any} theory of gravity with a covariant matter action with universal coupling.  The Riemannian Bianchi identity of the left-hand side of Einstein's equations implies only that the field equations provide no {\em additional} constraints on the motion beyond those arising from the matter action.  In other words, the field equations are {\em compatible} with the equations of motion.  This is also true in any alternative theory of gravity, such as scalar-tensor theory or Einstein {\AE}ther theory, provided that the theory does not contain what Thorne et al.\ \cite{1973PhRvD...7.3563T} call ``absolute elements'', non-dynamical fields such as a fixed flat background metric $\eta_{\mu\nu}$, or an absolute time parameter $T$. See \cite{1973PhRvD...7.3563T} for details.}.

We now wish to generalize this by including the possibility that the bodies that make up the material source may be self-gravitationally bound.  Because their structure may depend  on the ambient values of any auxiliary fields of the theory, the matter action will no longer be universally coupled, but will include dependence on those fields.  Accordingly, we now write the matter Lagrangian (see Eq. (\ref{eq:Lagrangian})) in the form
\begin{equation}
{\cal L}_M = {\cal L}_M (q_A, q_{A,\mu}, g_{\mu\nu}, g_{\mu\nu,\beta},Q_A, Q_{A,\mu}) \,,
\label{eq:Lagrangian2}
\end{equation}
where $Q_A$ represents the auxiliary fields.  Repeating the calculation shown above, and imposing the non-gravitational field equations (\ref{eq:nongrav}), we obtain the modified Bianchi identity
\begin{equation}
\nabla_\nu T^\nu_{\alpha} = (-g)^{-1/2} \sum_A \left [ Q_{A,\alpha} \frac{\delta \sqrt{-g}{\cal L}_M}{\delta Q_A} + 
\left ( D^\mu_{A\alpha} \frac{\delta \sqrt{-g}{\cal L}_M}{\delta Q_A} \right )_{,\mu} \right ] \,.
\label{eq3:bianchi3}
\end{equation}
where $D^\mu_{A\alpha}$ is the analogue of $d^\mu_{A\alpha}$  in Eq.\ (\ref{eq3:changeq}).

We consider scalar, vector and tensor auxiliary fields, and note that $D(\phi)^\mu_{\alpha} = 0$, $D(K^\nu)^{\mu}_{ \alpha} = K^\mu \delta^\nu_\alpha$, and 
$D(C^{\gamma\delta})^{\mu}_{ \alpha} = C^{\mu\delta} \delta^\gamma_\alpha
+ C^{\gamma\mu}\delta^\delta_\alpha$.  We then define the ``energy-momentum'' quantities
\begin{align}
T^{(\phi)} & \equiv \frac{1}{\sqrt{-g}} \frac{\delta \sqrt{-g}{\cal L}_M}{\delta \phi} \,,
\nonumber \\
T^{(K)}_\mu & \equiv -\frac{1}{\sqrt{-g}} \frac{\delta \sqrt{-g}{\cal L}_M}{\delta K^\mu} \,,
\nonumber \\
T^{(C)}_{\mu\nu} & \equiv \frac{2}{\sqrt{-g}} \frac{\delta \sqrt{-g}{\cal L}_M}{\delta C^{\mu\nu}} \,,
\label{eq:enmom}
\end{align}
(note that $T^{(C)}_{\mu\nu}$ need not be symmetric if $C^{\mu\nu}$ is not).
We substitute these relations into Eq.\ (\ref{eq3:bianchi3}), convert the various partial derivatives into covariant derivatives, demonstrate that all quantities involving Christoffel symbols vanish, and arrive finally at the covariant modified Bianchi identity
\begin{align}
\nabla_\nu \left [T^\nu_{\alpha} + K^\nu T^{(K)}_\alpha - \frac{1}{2} \left ( C^{\nu\delta} T^{(C)}_{\alpha\delta} + C^{\delta\nu} T^{(C)}_{\delta\alpha} \right ) \right ]  
&= \phi_{,\alpha}\, T^{(\phi)} - ( \nabla_\alpha K^\nu ) \, T^{(K)}_\nu+ 
\frac{1}{2} (\nabla_\alpha C^{\gamma\delta}) \,T^{(C)}_{\gamma\delta}   \,. 
\label{eq3:bianchi4}
\end{align}

We reiterate that the fundamental equation of motion in this class of metric theories is still 
$\nabla_\nu T^\nu_\alpha = 0$;  the extra terms in Eq.\ (\ref{eq3:bianchi4}) arise from attempting to account, in an effective field-theory manner, for the effect of auxiliary gravitational fields on the motion of {\em gravitationally self-bound} bodies, whose structure could be influenced by those fields.

As we have already discussed, one way to do this is to invoke the Eardley prescription, whereby we assume that the system consists of a set of ``point'' particles, whose masses depend on the local values of invariant quantities constructed from the auxiliary fields, with an action given by Eq.\ (\ref{eq:actionpoint}).  This can be converted into a spacetime action using delta functions; inserting $\int d^4x \delta^4 (x^\gamma - x_A^\gamma(\lambda) ) =1$ into Eq.\ (\ref{eq:actionpoint}) and expressing the action in the form of Eq.\ (\ref{eq:action}), we obtain
\begin{equation}
\sqrt{-g} {\cal L}_M = - \sum_A \int  m_A (\psi) (-g_{\alpha\beta} v^\alpha v^\beta )^{1/2}  \delta^4 (x^\gamma - x_A^\gamma(\lambda) ) \, d\lambda \,.
\label{eq:Lagrangian3}
\end{equation}
The energy momentum tensor is given by
\begin{equation}
T^{\mu\nu} = \frac{1}{\sqrt{-g}} \sum_A \int d\tau_A  \left [ \left (m_A (\psi) + m'_A  \frac{\partial \psi}{\partial u_A^\alpha} u_A^\alpha \right ) u_A^\mu u_A^\nu  - 2 m'_A \frac{\partial \psi}{\partial g_{\mu\nu}}  \right ]\delta^4 (x^\gamma - x_A^\gamma(\tau_A) ) \,.
\end{equation}
The other energy-momentum quantities can be obtained from Eqs.\ (\ref{eq:enmom}) and
(\ref{eq:Lagrangian3}). Equation (\ref{eq3:bianchi4}) then takes the form
\begin{align}
\nabla_\nu &\biggl [ \frac{1}{\sqrt{-g}} \sum_A \int d\tau_A \delta^4  (x^\gamma - x_A^\gamma(\tau_A) ) \times
\nonumber \\
&
\qquad 
\left \{
\left ( m_A  + m'_A  \frac{\partial \psi}{\partial u_A^\mu} u_A^\mu  \right ) u_A^\nu u_{A\alpha}  -  m'_A  \left ( 2 \frac{\partial \psi}{\partial g_{\mu\nu}} g_{\mu\alpha} -  \frac{\partial \psi}{\partial K^\alpha} K^\nu -  \frac{\partial \psi}{\partial C^{\alpha\delta}} C^{\nu\delta} 
-  \frac{\partial \psi}{\partial C^{\delta\alpha}} C^{\delta\nu} \right )
\right \}
\biggr ] 
\nonumber \\
&=
 -\frac{1}{\sqrt{-g}} \sum_A \int d\tau_A m'_A \delta^4  (x^\gamma - x_A^\gamma(\tau_A) )
 \left (  \frac{\partial \psi}{\partial \phi} \phi_{,\alpha} + 
 \frac{\partial \psi}{\partial K^\mu} \nabla_\alpha K^\mu
 + \frac{\partial \psi}{\partial C^{\gamma\delta}} \nabla_\alpha C^{\gamma\delta}
 \right ) \,.
 \label{eq3:bianchi5}
\end{align}

We now examine the form of these equations for each of the invariants described in Sec.\ \ref{sec:pointmass}.  For the invariants in category (A), $\partial \psi/\partial u_A^\mu = 0$, and it turns out that the quantities within the second parentheses on the left-hand side of Eq.\ (\ref{eq3:bianchi5}) cancel.  We are then left on the left-hand side with
\begin{align}
\nabla_\nu &\biggl [ \frac{1}{\sqrt{-g}} \sum_A \int d\tau_A m_A (\psi) u_{A\alpha} u_A^\nu \delta^4  (x^\gamma - x_A^\gamma(\tau_A) ) \biggr ]  
\nonumber \\
& \quad =  \frac{1}{\sqrt{-g}}  \sum_A \int d\tau_A m_A (\psi) \left [u_{A\alpha} u_A^\nu \partial_\nu \delta^4  (x^\gamma - x_A^\gamma(\tau_A) ) - \Gamma^\mu_{\alpha\nu} u_{A\mu} u_A^\nu \delta^4  (x^\gamma - x_A^\gamma(\tau_A) ) \right ]
\nonumber \\
& \quad =  \frac{1}{\sqrt{-g}}  \sum_A \int d\tau_A u_A^\nu \nabla_\nu [m_A (\psi) u_{A\alpha} ]\delta^4  (x^\gamma - x_A^\gamma(\tau_A) ) \,.
\label{eq:lhs1} 
\end{align}
\end{widetext}
We have used the fact that the partial derivative acts only on the free $x^\gamma$ in the delta function.  Then 
$u_A^\nu \partial_\nu \delta^4  (x^\gamma - x_A^\gamma(\tau_A) ) = -(d/d\tau_A) \delta^4  (x^\gamma - x_A^\gamma(\tau_A) )$; integrating by parts over $\tau_A$ and combining the result with the Christoffel symbol term yields the final result in Eq.\ (\ref{eq:lhs1}) (see \cite{PW2014} for details).
On the right-hand side, for invariants in this category, each term reduces to a partial derivative of the invariant itself.  The modified geodesic equations then have the same form as Eq.\ (\ref{eq:modgeodesicA}).

Turning to the invariants in category (B), it is simple to show in each case that the quantity inside the second parentheses on the left-hand side of Eq.\ (\ref{eq3:bianchi5})  has the form $B_\alpha u_A^\nu$, where  $B_\alpha = -K_\alpha$, $2C_{(\alpha\delta)} u_A^\delta$, $2C_{\epsilon\alpha} C^{\epsilon\delta} u_{A\delta}$ and  $2C_{\epsilon\alpha} C^{\delta\epsilon} u_{A\delta}$, respectively.  As in Eq.\ (\ref{eq:lhs1}) we convert 
these additional contributions to the left-hand side to the form  $(-g)^{-1/2} \sum_A \int d\tau_A u_A^\nu \nabla_\nu [m'_A B_\alpha ] \delta^4  (x^\gamma - x_A^\gamma(\tau_A) )$.
On the right-hand side of Eq.\ (\ref{eq3:bianchi5}), for each invariant, the result matches precisely the right-hand sides of Eqs.\ (\ref{eq:case1}) to (\ref{eq:case3}).

\section{Conclusions}
\label{sec:conclusions}

Using Eardley's prescription for incorporating the effects of self-gravity within gravitationally bound bodies in a range of alternative theories of gravity, we have derived exact, modified geodesic equations for a system of interacting bodies.  The equations are expressed in terms of the masses $m_A(\psi)$ and their derivatives $dm_A(\psi)/d\psi$, where $\psi$ is an invariant quantity constructed from the auxiliary gravitational field of the theory in question, whose value in the vicinity of each body affects its internal structure and thereby its mass.  The forms of the equations are independent of the field equations of the chosen theory.   These equations may be useful tools for deriving equations of motion for compact bodies to high post-Newtonian orders in alternative theories of gravity.

We have confined our attention to non-spinning bodies.  For spinning bodies, there could be additional couplings to the auxiliary fields, involving  invariants such as $u_{A\mu} K_\nu S^{\mu\nu}$, $C_{\mu\nu} S^{\mu\nu}$, and so on, where $S^{\mu\nu}$ is the spin tensor.   The methods of this paper can be applied to these cases straightforwardly, using well known delta-function methods adapted to spins.

\acknowledgments

This work was supported in part by the National Science Foundation,
Grant No.\ PHY 19-09247.   We are grateful for the hospitality of  the Institut d'Astrophysique de Paris where part of this work was carried out.


\end{document}